\documentclass[11pt]{article}

\def \lf  {\left (}
\def \rt  {\right )}
\def \be {\begin{equation}}
\def \ee {\end{equation}}
\def \a {\alpha}
\def \mn {\mu \nu}

\def \pl {\partial}

\begin{document}

\thispagestyle{empty}

\begin{center}

\bigskip

{\Large \bf Two Roads to the Null Energy Condition}

\bigskip
\bigskip
\bigskip

{\large \sc Maulik~Parikh\footnote{maulik.parikh@asu.edu}}

{\em Department of Physics \& Beyond Center for Fundamental Concepts in Science\\
Arizona State University, Tempe, Arizona 85287} 

\vspace*{1.5cm}

\large{\bf Abstract}
\end{center}
\noindent
The null energy condition has sweeping consequences in general relativity. I argue here that it has been misunderstood as a property exclusively of matter, when in fact it arises only in a theory of both matter and gravity. I then derive an equivalent geometric formulation of the null energy condition from worldsheet string theory, where it arises beautifully as simply Einstein's equations in two dimensions. But further, I show that this condition also has a thermodynamic origin, following from a local version of the second law of thermodynamics, applied to gravitational entropy. Thus, far from being an incidental property of matter, the validity of the null energy condition hints at the deep dual origins of gravity.
\bigskip
\bigskip
\bigskip


\bigskip
\bigskip
\bigskip


\newpage

\setcounter{page}{1}

\noindent
In 1949, upon hearing of G\"odel's solution of a universe with closed causal curves, Einstein remarked
\begin{quote}
cosmological solutions [...] have been found by Mr. G\"odel. It will be interesting to weigh whether these are not to be excluded on physical grounds. \cite{einstein}
\end{quote}
As Einstein recognized, general relativity needs to be supplemented by additional ``physical grounds" that might be used to exclude otherwise exact solutions of Einstein's equations. In  subsequent years, many plausible physical constraints have been proposed, from cosmic censorship to global hyperbolicity, but perhaps the most important -- because they are local -- are the energy conditions. 
It is difficult to overstate the significance of the energy conditions for general relativity. Energy conditions are invoked in the singularity theorems \cite{Penrose:1965}, in the topological censorship theorem \cite{topcensor}, in positive energy theorems, in the black hole no-hair theorem, in the laws of black hole mechanics \cite{Bardeen:1973gs}, as well as in excluding bouncing cosmologies \cite{molinaparis,nobounce}, traversable wormholes, the construction of time machines \cite{Hawking:1992}, and the creation of universes in a lab \cite{Farhi:1987}; in short, energy conditions support a host of gravitational theorems while protecting general relativity from the wilder fantasies of science fiction. The most fundamental of them, the null energy condition, requires that, at every point in spacetime,
\be
T_{\mn} v^\mu v^\nu \geq 0 \; ,	\label{TNEC}
\ee
for any light-like vector, $v^\mu$.

But what ``physical grounds" does the null energy condition flow from? In (\ref{TNEC}), $T_{\mn}$ is the classical energy-momentum tensor of matter. This suggests that (\ref{TNEC}) should arise, perhaps in some suitable classical limit, from some basic principle of quantum field theory, our best framework for describing matter. 
But quantum field theory does not appear to have a consistency requirement of the form of (\ref{TNEC}); indeed, we now know of several explicit examples of effective field theories, from ghost condensates to galileons, that violate (\ref{TNEC}) but that are nevertheless not in manifest conflict with quantum field theory. Moreover, (\ref{TNEC}) can be violated simply by switching the overall sign of the action, but the overall sign of the action is unphysical \cite{ArkaniHamed:2003}, being of no consequence in both classical and quantum field theory.
 

An important clue on how to proceed comes from the fact that the null energy condition is relevant primarily for theories in which matter is coupled to gravity \cite{necviolating}.
In such theories, Einstein's equations imply that the null energy condition can be reformulated in a quite different, though equivalent, form as
\be
R_{\mn} v^\mu v^\nu \geq 0 \; ,	\label{RNEC}
\ee
where $R_{\mn}$ is the Ricci tensor. Indeed, for the purpose of most gravitational theorems, the whole point of assuming (\ref{TNEC}) is to transmit the inequality to the Ricci tensor, so that it can be put to use in Raychaudhuri's equation. Our first key insight, then, is to regard (\ref{RNEC}), known as the Ricci or null convergence condition, as the actual physical requirement.
Note that, expressed in this form, we have a constraint on spacetime geometry, rather than energy densities. But the origin of such a condition is equally mysterious.



So now we are stuck with either a condition on matter, (\ref{TNEC}), which we cannot derive from quantum field theory, or a condition on geometry, (\ref{RNEC}), which we cannot derive from relativity.
But, as already suggested, if we are to derive (\ref{TNEC}) or (\ref{RNEC}) from first principles, we should do so in a theory that combines {\em both} matter and gravity. What theories, then, are neither general relativity nor quantum field theory and describe both matter and gravity? There are two that immediately spring to mind: string theory and thermodynamics.

String theory 
is particularly promising because it contains supergravity which, like most well-behaved theories, happens to satisfy the null energy condition as an unexplained fact. 
Consider then a bosonic closed string propagating in flat space. 
The Polyakov action is
\be
S[X^\mu,h_{ab}] = - \frac{1}{4 \pi \a'} \int d^2 \sigma \sqrt{-h} h^{ab} \pl_a X^\mu \pl_b X^\nu \eta_{\mn} - \frac{c}{4 \pi} \int d^2 \sigma \sqrt{-h}  R_h \; .  \label{flat-action}
\ee
Here $h_{ab}$ is the dimensionless worldsheet metric and $R_h$ is the corresponding Ricci scalar. 
The action is manifestly diffeomorphism- and Weyl-invariant. 
As the Lagrangian indicates, worldsheet bosonic string theory is simply a two-dimensional field theory of $D$ massless scalars, $X^\mu$, coupled to two-dimensional Einstein gravity. The equations of motion for $h_{ab}$, known as the Virasoro constraints, are just Einstein's equations in two dimensions; they are constraints because two-dimensional gravity is not dynamical. Yet the presence of $h_{ab}$ is necessary for worldsheet diffeomorphism invariance.

When the string propagates in curved space, the Minkowski metric $\eta_{\mn}$ is replaced by the general spacetime metric $g_{\mn}(X)$. We can perform a background field expansion $X^\mu(\tau, \sigma) = X^\mu_0 (\tau, \sigma) + Y^\mu (\tau, \sigma)$ where $X^\mu_0 (\tau, \sigma)$ is some solution of the classical equation of motion. Spacetime curvature turns (\ref{flat-action}) into an interacting theory. The resultant divergences can be cancelled by adding suitable counter-terms to the original Lagrangian. 
When the dust settles, and after integrating out $Y^\mu$, standard techniques 
\cite{CallanThorlacius,GSW1} yield the one-loop effective action:
\be
S[X_0^\mu,h_{ab}] = - \frac{1}{4 \pi \a'} \int d^2 \sigma \sqrt{-h} h^{ab} \pl_a X_0^\mu \pl_b X_0^\nu (\eta_{\mn} + C_\epsilon \a' R_{\mn} (X_0)) -  \frac{1}{4 \pi} \int d^2 \sigma \sqrt{-h} C_\epsilon \Phi(X_0) R_h \; . \label{eff-action}
\ee
Here $C_\epsilon$ is the divergent coefficient of the counter-terms. Consistent with symmetries, we have allowed there to be a dilaton field, $\Phi(X(\tau,\sigma))$. 

From this action, it is easy to work out the Virasoro constraints, i.e. Einstein's equations for the worldsheet metric, $h_{ab}$. In light-cone coordinates, these read
\be
0 =  \pl_+ X_0^\mu  \pl_+ X_0^\nu \left ( \eta_{\mu \nu} + C_\epsilon \alpha' R_{\mu \nu} + 2 C_\epsilon\alpha' \nabla_\mu \nabla_\nu \Phi \right ) \; , \label{curvedVirasoro}
\ee
and a similar equation with $+$ replaced by $-$.
To lowest order in $\a'$, we find
\be
\pl_+ X_0^\mu \, \pl_+ X_0^\nu \, \eta_{\mn} = 0 \; .
\ee
This is a very interesting equation. It says that the collection of $D$ scalar fields $X_0^\mu$ living on the worldsheet must collectively have the property that the vector $v^\mu = \pl_+ X_0^\mu$ is a {\em spacetime} null vector. We see that worldsheet string theory naturally singles out spacetime null vectors, a good omen if we are to derive the null energy condition.

Now consider  an arbitrary null vector $v^\mu$ in the tangent plane of some arbitrary point in spacetime. Let there be a test string passing through the given point with  $\pl_+ X_0^\mu$  equal to $v^\mu$ at the point. 
Then, at next order in $\a'$, (\ref{curvedVirasoro}) says
\be
v^\mu v^\nu (R_{\mn} + 2 \nabla_\mu \nabla_\nu \Phi  ) = 0 \; .
\ee
This is tantalizingly close to our form of the null energy condition, (\ref{RNEC}), but for two pesky differences: it is not an inequality, and there is an extra term involving the dilaton. 

Fortunately, we remember that the metric that appears in the worldsheet action is the string-frame metric. To transform to Einstein frame, define $g_{\mn} = e^{\frac{4 \Phi}{D-2}} g_{\mn}^E$. Then the Einstein-frame Ricci tensor satisfies
\be
R^E_{\mn} v^\mu v^\nu  = + \frac{4}{D-2} (v^\mu \nabla^E_\mu \Phi )^2 \; .
\ee
But the right-hand side is manifestly non-negative! Hence we have
\be
R^E_{\mn} v^\mu v^\nu  \geq 0 \; .
\ee
We see, remarkably, that the Virasoro constraints yield {\em precisely} the geometric form of the null energy condition \cite{NECderivation}, with string theory even supplying the contractions with null vectors.

We have thus found a very satisfying origin of the null energy condition. As anticipated, the first-principles origin of the null energy condition lies not in the quantum field theory of matter, nor in general relativity, but in string theory, which of course is a theory of both matter and gravity. 
Furthermore, we identify the ``physical grounds" behind the null energy condition as worldsheet diffeomorphism invariance (which leads to the Virasoro constraint); note that we did not explicitly invoke conformal invariance anywhere. It is particularly pleasing that a spacetime constraint is obtained from the Virasoro constraint which is itself a gravitational equation -- of two-dimensional gravity. Thus Einstein's equations in two dimensions restrict the physical solutions of Einstein's equations in spacetime, another example of the beautiful interplay between equations on the worldsheet and in spacetime. 

Although we assumed critical dimensionality, it can be shown that (\ref{RNEC}) also holds in lower dimensions, even after compactification has generated many more fields \cite{NECderivation}. Nevertheless, one critique of this derivation is that it seems tied to the properties of the field content of the low-energy limit of string theory. If the validity of the null energy condition is a property of all physical spacetimes, there ought to be a more manifest way
to see its universality. 

Indeed, there is, provided we regard gravity as emergent.

The notion of emergent gravity dates back at least to the work of Sakharov  \cite{Sakharov}. A more modern perspective, inspired by black holes, is that the microscopic theory lives in a lower dimension \cite{Verlinde}. 
Evidence for the emergence of gravity comes notably from the AdS/CFT correspondence in which gravity has a description in terms of a dual theory which does not itself contain gravity. From the perspective of the dual theory, gravity is an emergent phenomenon; the emergence of gravity is, in this sense, not a very controversial idea.

To obtain the null energy condition, we will appeal to the ultimate universal theory: thermodynamics. A relation between thermodynamics and the null energy condition is already present in black hole physics. Recall the derivation of the second law of thermodynamics for black holes \cite{Bardeen:1973gs}. The logic runs like this:
\be
T_{\mn} v^\mu v^\nu \geq 0 \Rightarrow R_{\mn} v^\mu v^\nu \geq 0 \Rightarrow \frac{d \theta}{d \lambda} \leq 0 \Rightarrow \theta \geq 0 \Rightarrow \frac{d A}{d \lambda} \geq 0 \Rightarrow \frac{d S}{d \lambda} \geq 0 
\ee
Here $\theta = \frac{1}{dA} \frac{dA}{d\lambda}$ is the expansion of a pencil of null generators of a black hole event horizon and $\lambda$ is a parameter along the null generators, which can be thought of as time. The first arrow follows from Einstein's equations, the second from the Raychaudhuri equation, the third from avoidance of horizon caustics, the fourth from the definition of $\theta$, and the last from the definition of Bekenstein-Hawking entropy. Ideally, we would like to reverse all these arrows so that the null energy condition flows from the second law of thermodynamics \cite{jordan}. But that looks like quite a tall order. On the right, the second law is a global statement -- one inequality per spacetime -- whereas on the left, the null energy condition is a local statement -- one inequality per spacetime {\em point}. Reversing the arrows appears impossible.



However, in a landmark paper \cite{einsteineqnofstate}, Jacobson was able to obtain Einstein's equations, which are also local, from essentially the first law of black hole thermodynamics. The key idea was to assume that, in keeping with the universality of horizon entropy, the first law could be applied to local Rindler horizons. Thus a global law was ``gauged," which is a pre-requisite for obtaining local equations of motion. In the same vein, we will show that even the null energy condition, in the form of the Ricci convergence condition, (\ref{RNEC}), comes out of thermodynamics applied to some covariantly defined null holographic screen \cite{Bousso}. In a nutshell, just as Jacobson regarded the first law as an input and obtained Einstein's equations as an output (reversing the laws of black hole mechanics, as it were), we shall regard the second law as an input and obtain the null energy condition as an output \cite{thermoNEC}.

Consider then an underlying non-gravitational system with a finite number of degrees of freedom. This means it can have a maximum coarse-grained entropy, $S_{\rm max}$. Assume that the theory obeys the second law of thermodynamics. Incredibly, from these mildest of assumptions and one more, we can derive the null energy condition. As the system relaxes to its equilibrium, its coarse-grained entropy at late times generically takes the form
\be
S(t) \approx S_{\rm max} (1- e^{-k t}) \; , \label{relax}
\ee
where $k^{-1}$ is the relaxation time. Then we have
\be
\ddot{S} \leq 0 \label{ddot}
\ee
This is the result we will need. Any system with finite entropy and obeying the second law must, on time-scales over which transient fluctuations are smoothed out, obey (\ref{ddot}) as it approaches equilibrium. In fact, 
it can be shown quite generally \cite{entropyproduction} that near-equilibrium systems approaching equilibrium obey $\ddot{S} \leq 0$, with saturation attained at equilibrium. 

Now comes the last assumption. We assume that the coarse-grained entropy corresponds to the area of the holographic screen of some infinitesimal region of spacetime. Precisely the same assumption was made by Jacobson. 
We then have
\be
S = \frac{A}{4} \; ,
\ee
so that
\be
\dot{S} = \frac{A}{4} \theta \; .
\ee
Here $\theta$ can be regarded as constant over the surface since the screen is infinitesimal. Then
\be
\ddot{S} = \frac{A}{4} \lf \theta^2 + \dot{\theta} \rt \; .
\ee
Since holographic screens are null, they obey the optical Raychaudhuri equation. Applying that equation to the null generators of the screen (and neglecting shear and vorticity terms) we find \cite{thermoNEC}:
\begin{eqnarray}
R_{\mu \nu} v^\mu v^\nu & = & - \dot{\theta} - \frac{1}{2} \theta^2 \nonumber \\
& = & - ( \dot{\theta} + \theta^2 ) + \frac{1}{2} \theta^2 \nonumber \\
& = &  - \frac{\ddot{S}}{S} + \frac{\dot{S}^2}{2S^2}  \nonumber \\
& \geq & 0 \; .
\end{eqnarray}
The origin of the geometric null energy condition is the second law of thermodynamics!

In fact, a second-law origin of the null energy condition tells us more. We know that the second law can be violated by fluctuations. The rate of a fluctuation to some macrostate with entropy $S_f$ is
\be
\Gamma \sim e^{S_f - S_{\rm max}} = e^{\Delta S} \; .
\ee
But we know this formula very well: it is the tunneling rate \cite{secret,energyhawking,tunneling} of Hawking radiation in a theory of quantum gravity. A violation of the null energy condition can thus be viewed as a {\em quantum} effect in gravity but a {\em thermal} effect in the microscopic theory.

Finally, we ask: why are there two independent derivations of the null energy condition? Actually, there are also two derivations of Einstein's equations. In string theory, Einstein's equations arise from demanding Weyl invariance on the worldsheet. But in emergent gravity, they arise from the first law of thermodynamics applied to local Rindler horizons. Here we have seen that the null energy condition is also built on dual ``physical grounds": it arises out of worldsheet diffeomorphism invariance and emerges from the second law of thermodynamics. This can be summarized in an extraordinary table:
\vspace{3mm}

\noindent
\begin{tabular}{r | c | l}
{\bf String Worldsheet} & {\bf Spacetime} & {\bf Thermodynamics} \\ \hline \hline
Weyl invariance $\Rightarrow$ & Einstein's equations & $\Leftarrow$ first law \\
diffeomorphism invariance $\Rightarrow$ & null energy condition & $\Leftarrow$ second law
\end{tabular}

\vspace{3mm}
\noindent
Cover up the central column. Is there a relation between the first and third columns? Are the dual origins just a miraculous coincidence? As our hero once said with regard to another supposed coincidence, ``Subtle is the Lord, but malicious He is not."

\end{document}